\documentstyle[12pt]{article}

\newcommand{\eq}{\begin{equation}}
\newcommand{\en}{\end{equation}}

\newcommand{\NP}[1]{Nucl.\ Phys.\ {\bf #1}}
\newcommand{\PL}[1]{Phys.\ Lett.\ {\bf #1}}

\newcommand{\PR}[1]{Phys.\ Rev.\ {\bf #1}}

\begin{document}

\hskip 12cm \vbox{\hbox{DFTT 41/91}\hbox{September 1991}}
\vskip 0.4cm
\centerline{\bf   THE EFFECTIVE STRING OF 3D  GAUGE SYSTEMS}
\centerline{\bf AT THE DECONFINING TRANSITION$^\star$}
\vskip 1.3cm
\centerline{ M. Caselle$^{\diamond}$  and  F. Gliozzi}
\vskip .6cm
\centerline{\sl  Dipartimento di Fisica
Teorica dell'Universit\`a di Torino}
\centerline{\sl Istituto Nazionale di Fisica Nucleare,Sezione di Torino}
\centerline{\sl via P.Giuria 1, I-10125 Turin,Italy}
\vskip 1.5cm
\vskip 1.5cm

\begin{abstract}
It is argued that the effective string of whatever 3D gauge system at
the deconfining transition is universally described by the minimal $N=2$
extended superconformal theory at $c=1$. A universal value of the
critical temperature is predicted.
\end{abstract}
\vskip 2cm
\hrule
\vskip1.2cm
\noindent

\noindent
$^{\ast}${\it Work supported in part by Ministero dell'Universit\`a e della
Ricerca Scientifica e Tecnologica}

\hbox{\vbox{\hbox{$^{\diamond}${\it email address:}}\hbox{}}
 \vbox{\hbox{ Decnet=(39163::CASELLE)}
\hbox{ Bitnet=(CASELLE@TORINO.INFN.IT)}}}
\vfill
\eject

\newpage
\setcounter{page}{1}
Consider a pair of quarks propagating in a gauge medium at a temperature
$T$ below the deconfining point $T_c$. This system is described by the
static potential
\eq
V(R)= \sigma(T)\, R \,-\,\kappa\frac{\pi}{R}\,+\,O(1/R^2)~~~,
\label{V}
\en
where $\sigma$ is the tension of the effective string describing the
infrared behaviour of the theory~\cite{NO} and $\kappa\pi/R$ is the
universal, long distance, Coulomb term generated by the the zero point
energy of the effective string~\cite{{brink},{luscher}}.

\vskip .3 cm
On very general grounds , the effective string theory can  be considered
as a conformal field theory (CFT) on a surface with quark lines as
boundaries. Recently, utilizing the analysis of Cardy~\cite{cardy} on
the effects of boundaries on CFT, the general form of the universal
constant $\kappa$ as a function of the conformal parameters has been
discussed~\cite{noi}
\eq
\kappa\,=\,\frac{c}{24}-h
\label{k}
\en
where the conformal anomaly $c$ measures the physical degrees of freedom
of the string and $h$ is the minimal conformal width of the primary
fields compatible with the boundary conditions at the ends of the string.

Simple physical properties of the colour flux tubes in the confined
phase suggest a very specific CFT ~\cite{noi} as a model of the string
at zero temperature, but here we need only to retain the
(generally accepted ) value of the conformal anomaly or central charge
$c$
\eq
c=D-2~~~,
\label{c}
\en
which encodes in the CFT the fact that at large distances , at least,
the only physical degrees of freedom of the effective string are the
transverse displacements of the world-sheet.

For $c\geq1$ there are, in general, in the spectrum of the theory,
marginal operators , allowing deformations of the CFT, preserving
the conformal symmetry and the central charge $c$. Typical examples
of the parameters, or moduli, controlling these deformations are the
compactification radii.  In a finite temperature gauge system the
displacements $x_\perp$ of the effective string in the direction of
of the imaginary time axis is actually compactified on a circle of
radius $L/2\pi=1/(2T\pi)$
\eq
x_\perp\sim x_\perp +L~~~,
\label{T}
\en
therefore the temperature $T$ is a conformal modulus of the CFT: varying
 $T$ modifies the spectrum of the theory, i. e. $h=h(T)$.

\vskip.3 cm
At the deconfining point $T=T_c$ the string tension $\sigma(T_c)$
vanishes and the flux  connecting the two quarks cannot longer be
described by an effective string. As  a consequence,  the long
distance Coulomb term looses its very justification: this $1/R$
behaviour  was produced by the quantum fluctuations of the
string, but now the string has disappeared . Consistency requires
vanishing of $\kappa$  at $T=T_c$, i.e.
\eq
h(T_c)=\frac{c}{24}~~~.
\label{hc}
\en

For $D=3$ (i.e. $c=1$ ) the moduli space is one-dimensional, so
eq.\ref{hc} uniquely fixes, as we shall see, not only  a specific CFT,
but also a value
for the deconfining temperature. Indeed , in terms of the adimensional
modulus $r$, defined by
\eq  r=\frac{\sqrt\sigma}{2\sqrt{\pi}T}~~~,
\label{r}
\en
the conformal width of the primary fields is  simply given
by~\cite{cremmer}
\eq
h=\frac{(mr+n/2r)^2}{2}~~~.
\label{spect}
\en
We are of course interested only to CFT with a discrete spectrum,
corresponding to $r^2$ rational.
Inserting eq.\ref{spect} into eq.\ref{hc} selects the following special
radii
\eq
r_1=\frac{1}{2\sqrt3}\;,\;r_2=\frac{1}{\sqrt3}\;,\;
r_3=\frac{\sqrt3}{2}\;,\;r_4=\sqrt3\;,
\label{rc}
\en
which exactly correspond to the only points where the conformal symmetry
is promoted to a $N=2$ extended supersymmetry~\cite{ademollo}. The
spectrum of the primary fields in these four points is that of the
 unitary $N=2$ minimal model at $c=1$~\cite{DPZ}.

\vskip.3 cm
Eq.\ref{rc} singles out , through eq.\ref{r}, four special temperatures.
 It is reasonable to assume that the deconfining temperature corresponds
to the minimal radius $r_1$, while the others correspond to metastable
solutions. This gives

\eq
\frac{T_c}{\sqrt\sigma}=\frac{\sqrt3}{\sqrt\pi}~~~.
\label{ng}
\en
Remarkably enough, it coincides with the value predicted
for the Nambu-Goto string~\cite{pisarski}. Our argument suggests that
this temperature is universal and independent on the gauge group.

\vskip.3 cm
The only existing data on the deconfinement temperature of (2+1) LGT
are, to our knowledge, for the SU(2) and $Z_2$ gauge groups.
Let us now compare them with our prediction.

\begin{description}
\item{SU(2)]}
 In this case there is a good convergence among different published
results both for the string tension and for the critical temperature.
Data on $T_c$ are reported in~\cite{tc1} for lattices with length $L=2$
in
the imaginary time direction; in~\cite{tc2} for $L=4$ and in~\cite{tc3}
for $L=5$ and 6. Scaling seems to be fulfilled already for $L>4$
and allows a rather precise estimation: $T_c\beta a=1.4\pm0.1
$~\cite{tc3} (where $a$ is the lattice spacing and $\beta$ the inverse
of the coupling constant).
Data on $\sigma$ are reported in~\cite{sigma1} and more
recently in~\cite{sigma2}, in both case obtained with Wilson loops.
Scaling seems to be fulfilled already for $\beta>6.$ (in agreement with
the $T_c$ behaviour). These two sets of data agree and
give $\sigma(\beta a)^2=2.23\pm0.04$.  This implies
$T_c/\sqrt{\sigma}=0.94\pm0.03$ in agreement with our critical
temperature for the minimal radius $r_1$,
$T_c/{\sqrt\sigma}={\sqrt3}/{\sqrt\pi}=0.98$.
 Indeed this result was
already presented in~\cite{fp} to support the Nambu-Goto string picture, and
has not been significatively changed by the new high precision data of
ref~\cite{sigma2}.

\item{$Z_2$]}
 In this case a rather precise estimate exists for the string
tension ~\cite{noiz2}~:
$\sigma a^2(\beta_c-\beta)^{-2\nu}=3.64\pm0.09$~, with
$\beta_c=0.7614$ and $\nu=0.63$.
For the critical
temperature, the scaling region seems to start from $L=4$. Data obtained
with the Montecarlo renormalization group approach~\cite{wan2} for
$L=4,8$, give $T_c a(\beta_c-\beta)^{-\nu}
=2.3\pm0.1$, while a slightly lower value was
obtained using finite size scaling techniques~\cite{wan1,brower}
for $L=4$:  $T_c a(\beta_c-\beta)^{-\nu}=2.2\pm0.1$. Taking
into account all the uncertainties we
can assume: $T_c/\sqrt{\sigma}=1.17\pm0.1$ which is in acceptable
agreement with our solution.

\end{description}
\vskip .3cm
It would be interesting to test eq. \ref{ng} with other gauge groups.

\vfil
\end{document}